\documentclass{article}
\usepackage{spconf,amsmath,graphicx,hyperref,amsfonts}
\usepackage{bm, bbm} 
\usepackage{multirow}
\usepackage{booktabs}
\usepackage[normalem]{ulem}

\title{Revise-TTS: A Non-Masking Refinement Framework for Masked Diffusion Models in Text-to-Speech}
\name{Wanli Sun$^{1, 2}$\sthanks{Work done during an internship at HojoAI.} \qquad  Dan Su$^{2}$ \qquad Anton Ragni$^{1}$}
\address{
$^{1}$ School of Computer Science, University of Sheffield, Sheffield, UK \qquad $^{2}$ HojoAI, Beijing, China
}

\begin{document}
\ninept
\maketitle
\begin{abstract}
Text-to-speech (TTS) based on masked diffusion models (MDMs) achieves competitive zero-shot speech synthesis performance. 
% with conventional TTS based on autoregressive (AR) token modeling. 
Unlike autoregressive (AR) models, MDMs define a corruption process that maps clean speech-token sequences to masked maps and learn a reverse denoising process to recover the reference tokens.
However, MDMs commit to intermediate predictions that cannot be revised in subsequent inference steps. Consequently, prediction errors accumulate and propagate, degrading synthesized speech quality.
We present Revise-TTS, a single-stage framework that applies hierarchical binary quantization (HBQ) and generative refinement networks (GRN) to zero-shot speech generation.
First, we replace finite scalar quantization (FSQ) with HBQ to introduce a scalable binary representation for speech tokens with an explicit coarse-to-fine structure.
% Secondly, rather than using fully masked initialization, the GRN framework is trained to predict codebook tokens from randomly initialized token sequences. 
Second, GRN is trained to reconstruct the token sequence from a mixture of ground-truth and random tokens. 
During inference, iterative refinement enables self-correction by randomly discarding a small proportion of previously predicted tokens with new predictions.
Experiments and ablation studies demonstrate that Revise-TTS achieves competitive zero-shot speech synthesis performance using HBQ. 
It can further be extended to any MDM-based TTS task as an alternative training and inference method.
% The abstract should appear at the top of the left-hand column of text, about
% 0.5 inch (12 mm) below the title area and no more than 3.125 inches (80 mm) in
% length.  Leave a 0.5 inch (12 mm) space between the end of the abstract and the
% beginning of the main text.  The abstract should contain about 100 to 150
% words, and should be identical to the abstract text submitted electronically
% along with the paper cover sheet.  All manuscripts must be in English, printed
% in black ink.
\end{abstract}
\section{Introduction}
\label{sec:intro}
Zero-shot text-to-speech (TTS) has recently achieved high-quality speech generation through autoregressive (AR)~\cite{xie2025fireredtts, gong2026moss} and non-autoregressive (NAR) modeling~\cite{wang2025maskgct, zhu2026omnivoice, chen2025f5, 11433841}. AR models generate tokens from left to right, causing error accumulation and requiring $N$ decoding steps for an $N$-token sequence. In contrast, NAR models generate all tokens in parallel and generally require far fewer inference steps, reducing latency and avoiding AR error accumulation. Masked diffusion models (MDMs)~\cite{austin2021structured, sahoo2024simple, Lou2023DiscreteDM} are an important NAR framework that predicts masked tokens over multiple denoising steps. However, typical MDM decoding commits non-overlapping indices across intermediate steps, preventing high-confidence but incorrect predictions from being revised. This problem is further complicated in two-stage TTS systems with separate text-to-semantic and semantic-to-acoustic modules, where the source of a generation error is difficult to identify. Recent advances in single-stage TTS~\cite{gallego2025single, zhu2026omnivoice, chen2026sac} suggest that revisable generation algorithms can be developed directly within the NAR TTS framework.

Despite this progress, discrete-token quantization in TTS remains limited. FSQ~\cite{mentzer2024finite} independently quantizes each latent dimension into fixed scalar levels, producing a flat representation, while higher precision requires more dimensions. Larger codebooks preserve finer acoustic details and reduce quantization error but expand the token prediction space and complicate training. Conversely, smaller codebooks are easier to optimize but may discard important linguistic and speaker information.

Error revision algorithms have been explored in visual synthesis~\cite{lezama2022discrete, han2026generative} and text generation~\cite{schiff2026learn, wang2026remasking, zhao2026informed}. These methods either improve sampling quality~\cite{wang2026remasking} or add modules to identify incorrect positions~\cite{lezama2022discrete} or tokens~\cite{zhao2026informed, schiff2026learn}. Our approach follows the Generative Refinement Networks (GRN) framework~\cite{han2026generative}, which treats generation as iterative global refinement rather than masked-token prediction. By repeatedly revising partial sequences, later steps can correct earlier errors. We also introduce Hierarchical Binary Quantization (HBQ) to address FSQ bottlenecks. HBQ represents continuous latent features through multi-level binary rounds, yielding a hierarchical representation with progressively higher precision in TTS.
To address the revision in MDM-based TTS, this paper investigates the use of overlapping in MDM-based TTS following GRN in visual synthesis. 
Concretely, it investigates utilizing GRN in TTS model and HBQ in the quantization of speech tokens.

% To address the revision in MDM-based TTS, this paper investigates the use of overlap in MDM following GRN in visual synthesis. Concretely, it investigates utilizing GRN in TTS model and HBQ in the quantization of speech tokens.
% Experiments show that HBQ outperforms other quantization algorithms while achieving strong performance with a relatively small codebook. 
% % GRN also uses uncertainty or entropy to allocate more computation to difficult samples, providing a natural path toward adaptive inference. 
% The results further demonstrate that GRN can serve with an overlap 
% to alleviate the revision for improving speech quality.

In summary, our contributions can be summarized as follows:
\begin{itemize}
    \item the first error-correction MDM-based TTS model;
    \item an exploration of HBQ for speech-token quantization;
    \item an extended framework for MDM-based TTS .
\end{itemize}
% These guidelines include complete descriptions of the fonts, spacing, and
% related information for producing your proceedings manuscripts. Please follow
% them and if you have any questions, direct them to Conference Management
% Services, Inc.: Phone +1-979-846-6800 or email
% to \\\texttt{papers@2027.ieeeicassp.org}.

\section{Methodology}
Masked Diffusion Models (MDM)~\cite{shi2024simplified, sahoo2024simple} are parametric models $p_{\bm \theta}$ trained to denoise a fixed noising process $q$, which maps a clean speech-token map \(\bm Y_0\) to a progressively masked map \(\bm Y_t\) at noise level \(t\in[0,1]\).
% that maps clean speech tokens $\bm Y_0$ to increasingly noisy latent variable $z_t$, for t [0, 1]. 
Let $\bm Y_0 \in \mathcal V^{T \times Q}$ denote the acoustic token map of the target speech, where $T$ is the sequence length and $Q$ is the number of token codebooks. Following the simplified corruption process~\cite{sahoo2024simple}, also known as ``absorbing-state'' diffusion, the probability at each target position $(i,j)$ for any noise level $t$ is given by:
\begin{equation}
% \begin{aligned}
q(\bm{Y}_t \mid \bm{Y}_0, t)
=\prod_{i=1}^T \prod_{j=1}^Q \Bigl[t \cdot \delta_{\mathrm{M}}(y_t^{(i, j)})+(1-t) \cdot \delta_{y_0^{i, j}}(y_t^{(i, j)})\Bigr],
% \end{aligned}
\label{MDM_training}
\end{equation}
% where ${\delta_M}$ denotes a delta function applied on masking positions. 
where $\mathrm M$ denotes the mask token and $\delta_a(y)=\mathbbm{1}[y=a]$ denotes a point mass at $a$.
During denoising, the model predicts the most likely clean token at each masked position, conditioned on the input text $\bm x$, the optional speech prompt $\bm Y^P$, and the corrupted token map \(\bm{Y}_t\), as follows:
\begin{equation}
    \widehat{y}_0^{(i,j)} =\arg\max_{v\in\mathcal{V}} p_{\bm \theta}\!\left( y_0^{(i,j)}=v \mid \bm{Y}_t, t; \bm{x},\bm{Y}^{P} \right),
\label{eq:mdm_denoising}
\end{equation}
where \(\mathcal V\) denotes the vocabulary of acoustic tokens and \(v\in\mathcal V\) is a candidate token. At each denoising step $t$, only the highest-confidence predictions are committed to update $\bm Y_t$, which is then reused as the current token map in the following iteration. Confidence is defined as the maximum predicted token probability. The number of committed predictions is determined by a masking schedule function, e.g., a linear schedule~\cite{shi2024simplified} or a cosine schedule~\cite{zhu2026omnivoice}. The remaining positions stay masked, while previously committed tokens remain unchanged. Consequently, an incorrectly committed token cannot be revised and may influence subsequent predictions.

\subsection{Hierarchical binary quantization}
\label{sec:HBQ}
Recent audio tokenizers~\cite{bosonai_higgs_audio_tts_v3_2026, Ye2025LlasaST, borsos2023audiolm} follow an autoencoder architecture, 
where an acoustic encoder extracts acoustic features for reconstruction, while a semantic encoder extracts semantic features that capture linguistic informations.
% provide auxiliary supervision, encouraging the learned tokens to preserve not only acoustic representations but also linguistic information.
These two features are concatenated to form fused representations $\bm H\in\mathbb{R}^{T\times C}$, where $T$ is the token-sequence length and $C$ is the number of latent channels.
During training of audio tokenizers, the fused representations are quantized into discrete tokens, which are used for both acoustic reconstruction and auxiliary semantic supervision.
During inference, only the acoustic decoder is used to reconstruct speech from the quantized representations.
% using an acoustic encoder and a semantic encoder in parallel to extract acoustic and semantic features from a raw speech waveform ${\bm Y}$. 
% During inference, the speech waveform is reconstructed using only the acoustic features. 
% During training, the acoustic and semantic features are concatenated to form a sequence of fused representations ${\bm h_{1:T}}$ of length ${T}$.
% are projected to ${C}$ channels using a fully-connected layer. 
% Unlike visual synthesis methods that transform VAE features into tokens~\cite{han2026generative}, we transform fused representations into tokens. 
 
\begin{figure}[ht]
    \centering
    \makebox[\columnwidth][c]{\includegraphics[width=0.9\columnwidth]{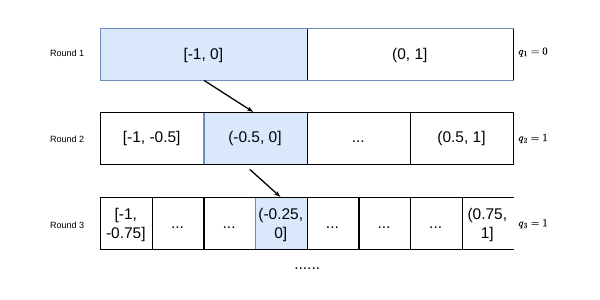}}
    \caption{Illustration of HBQ. Each round halves the selected quantization interval. For $z=-0.113$, three rounds produce the bit sequence $011$, corresponding to the discrete index $3$.}
    \label{fig:HBQ}
\end{figure}

Hierarchical binary quantization (HBQ)~\cite{han2026generative} is applied to quantize the fused representation $\bm H$.
% while avoiding an explicit VQ objective term during training, such as the commitment loss in residual vector quantization (RVQ). 
First, HBQ normalizes the fused representation element-wise as $\bar{\bm H}=\tanh(\bm H)$.
As illustrated in Figure~\ref{fig:HBQ}, HBQ performs $M$ successive binary refinement rounds for each scalar $z\in(-1,1)$ in $\bar{\bm H}$. At round $m$, the decision centre $c_m$ and binary label $b_m\in\{0,1\}$ are computed as follows:
\begin{align}
c_m &= \sum_{r=1}^{m-1}\frac{2b_r-1}{2^r}, \qquad c_1 = 0,\\
b_m &=
\begin{cases}
0, & z\leq c_m,\\
1, & z>c_m,
\end{cases}
\qquad m=1,\ldots,M,
\end{align}
% We apply hierarchical binary quantization
% (HBQ)~\cite{han2026generative} to the fused representation
% $\bm H\in\mathbb{R}^{T\times C}$. HBQ first normalizes it element-wise
% as $\bar{\bm H}=\tanh(\bm H)$. For each scalar
% $z\in(-1,1)$ in $\bar{\bm H}$, $M$ coarse-to-fine binary refinement
% rounds are performed:
% \begin{align}
% \mu_m
% &=
% \sum_{r=1}^{m-1}\frac{2b_r-1}{2^r},
% \qquad \mu_1=0, \label{eq:hbq_center}\\
% b_m
% &=
% \begin{cases}
% 0, & z\leq\mu_m,\\
% 1, & z>\mu_m,
% \end{cases}
% \qquad m=1,\ldots,M. \label{eq:hbq_bit}
% \end{align}
% The resulting labels $\{b_1,\ldots,b_M\}$ approximate $z$ by
% $\hat z=\sum_{m=1}^{M}(2b_m-1)/2^m$, with an error bounded by
% $2^{-M}$. They form $\bm Y_{\mathrm{bit}}$ directly or are packed
% into $\bm Y_{\mathrm{ind}}$. Accordingly, in
% Eqs.~\eqref{MDM_training}--\eqref{eq:mdm_denoising},
% $(Q,\mathcal V)=(CM,\{0,1\})$ for $\bm Y_{\mathrm{bit}}$ and
% $(Q,\mathcal V)=(C,\{0,\ldots,2^M-1\})$ for
% $\bm Y_{\mathrm{ind}}$.

% To obtain discrete tokens, we use HBQ to transform each normalized continuous value ${v}$ in each channel of codebook into binary labels as follows:
% \begin{align}
% c_i &= \sum_{j=1}^{i-1} \frac{\delta\left(q_j\right)}{2^j}, \\
% q_i &=
% \begin{cases}
% 0, & \text{if } v \leq c_i \\
% 1, & \text{if } v > c_i,
% \end{cases}
% \end{align}
% where ${c_1}=0$. ${\delta(\cdot)}$ denotes a delta function with -1 when $q_j=0$ and 1 otherwise. 
After $M$ refinement rounds, each normalized value ${z}$ is represented by the binary labels $\{b_1, b_2, \dots, b_M\}$. 
As illustrated in Figure~\ref{fig:HBQ}, $z=-0.113$ is encoded as the bit sequence $011$, which corresponds to the discrete index $3$ and the quantized value $\hat z=-0.125$.
Each refinement round halves the current interval. Thus, after $M$ rounds, each code represents an interval of width $2^{1-M}$ in the normalized latent space, yielding a maximum quantization error of $2^{-M}$. 
% Here each token performs a range of $2^{1-M}$ centred with ${c_M}$, which quantization error is less than $2^{M}$.
Encoding fused representations using a codebook of size $2^{CM}$ (e.g., $C=16$ and $M=4$) is challenging for TTS models because of the large codebook size.
Following prior work~\cite{han2026generative}, two compact representations are considered: $\bm Y_{\text{ind}} \in\left\{0, \ldots, 2^M-1\right\}^{[T, C]}$ and $\bm Y_{\text{bit}} \in\{0,1\}^{[T, C, M]}$. 
% we represent the binary labels $\{q_1, q_2, \dots, q_M\}$ in two alternative forms with smaller codebook sizes: $\bm Y_{\text {ind }} \in\left\{0, \ldots, 2^M-1\right\}^{[B, T, C]}$ or $\bm Y_{\text {bit}} \in\{0,1\}^{[B, T, C, M]}$. 
% Applying these two representations to TTS yields $\mathrm{GRN}_{\mathrm{ind}}$ and $\mathrm{GRN}_{\mathrm{bit}}$.
% In Eqs.~\eqref{MDM_training}--\eqref{eq:grn_objective},
% $\bm Y_0$ denotes either $\bm Y_{\mathrm{ind}}$, for which
% $(Q,\mathcal V)=(C,\{0,\ldots,2^M-1\})$, or the flattened
% $\bm Y_{\mathrm{bit}}$, for which $(Q,\mathcal V)=(CM,\{0,1\})$.
Both preserve the same information capacity of $CM$ bits per frame, but the use of $\bm Y_{\text {ind}}$ in the cascading TTS models would require $C2^M$-way classifications, whereas $\bm Y_{\text {bit}}$ performs $CM$ binary classifications.

HBQ and FSQ are both lookup-free quantization algorithms that avoid an explicit VQ objective term during training, such as the commitment loss used in vector quantization (VQ) or residual vector quantization (RVQ).
HBQ quantizes each feature dimension using an explicit coarse-to-fine ordering, whereas FSQ quantizes each feature dimension using a fixed set of scalar levels.
In addition, the outputs of HBQ can be used directly for bitwise prediction or packed into an integer index. The following experiments compare FSQ with HBQ configurations using different combinations of $C$ and $M$ to demonstrate the effectiveness of HBQ.
% \begin{equation}
% \hat{F}=\delta\left[q_1\right] \cdot 2^{-1}+\delta\left[q_2\right] \cdot 2^{-2}+\ldots+\delta\left[q_M\right] \cdot 2^{-M}
% \end{equation}
% \subsection{Generative refinement network: training}
\subsection{Generative refinement network for TTS}
Although HBQ provides a compact quantization algorithm for MDM-based TTS, improved quantization does not address the irreversible prediction errors during iterative decoding. Therefore, an overlapping decoding for MDM following GRN~\cite{han2026grn} is adopted, enabling the model to preserve correct tokens while replacing incorrect tokens in subsequent iterations.

Using the two HBQ representations defined above lead to two modes: $\mathrm{GRN}_{\mathrm{ind}}$ and $\mathrm{GRN}_{\mathrm{bit}}$,which are in the index and flattened bit token spaces, respectively. Both variants share the training and sampling procedures described below.
% Given the target token map ${\bm Y_t}$ a time step $t$ and a random token map ${\bm Y_{rand}}$, the objective is to predict the next step target token map ${\bm Y_{t+1}}$. Therefore, we randomly combine the target token map and a random token map to imitate the optimizing objective token map $\bm F_t$. 
% During training, we do not explicitly sample a discrete time step $t$. Instead, we uniformly sample a retention ratio $l\in[0,1]$ and construct the corrupted acoustic-token sequence $\bm{F}$ according to Eq.~\ref{eq:grn_training_corruption}. 
% \subsubsection{Training}
\paragraph*{Training}
\label{sec:grn_training}
At each training iteration, we independently sample a retention ratio $l\sim\mathcal{U}(0,1)$. 
A small $l$ produces a heavy corruption, whereas a large $l$ produces a light corruption. 
Uniform sampling therefore exposes a single model to the continous corruption levels encountered during iterative refinement.
A binary selection map $\bm{S}$ is then sampled according to  $l$ and used to construct the corrupted input token map $\bm F$ by mixing the ground-truth token map $\bm{Y_0}$ with uniformly sampled random tokens:
% \begin{equation}
% \bm F= \bm S \cdot \bm Y + (1 - \bm S) \cdot \bm Y_{\mathrm {rand}},
% \end{equation}
\begin{equation}
\bm F = \bm S\odot\bm Y_0 + (\bm 1-\bm S)\odot\bm Y_{\mathrm{rand}},
\end{equation}
where $\bm Y_{\mathrm{rand}}$ is a randomly sampled token map with the same shape as $\bm Y_0$. 
Given conditioning inputs such as text ${\bm x}$ and prompt speech ${\bm Y^P}$, the model is trained to reconstruct all target acoustic tokens using the cross-entropy loss:
% \begin{equation}
% \mathcal{L}_{\mathrm{GRN}}
% =
% -\frac{1}{N}
% \sum_{n=1}^{N}
% \log p_{\bm \theta}
% \left(
% y_n \mid \bm{F},\bm{x},\bm{Y}^{P}
% \right),
% \label{eq:grn_objective}
% \end{equation}
\begin{equation}
\mathcal{L}_{\mathrm{GRN}}
= -\frac{1}{TQ}\sum_{i=1}^{T}\sum_{j=1}^{Q} \log p_{\bm\theta}\!\left( y_0^{(i,j)} \mid \bm F,\bm x,\bm Y^P \right).
\label{eq:grn_objective}
\end{equation}
% where ${N}$ is the length of the flattened ground-truth token sequence. 
The objective in Eq.~\ref{eq:grn_objective} trains GRN to reconstruct the entire target token map from partially corrupted inputs, thereby replacing corrupted tokens while preserving correct ones. During inference, the retention ratio follows a cosine schedule function and no separate policy network is learned for token selection.
% Conventional MDMs learn to predict masked tokens and provide confidence scores for their predictions, but the selection of tokens to retain is typically governed by a predefined confidence-based remasking strategy rather than an explicitly learned retention policy.
% With the objective in Eq.~\ref{eq:grn_objective}, GRN is trained to correct corrupted tokens while preserving tokens that are already correct. Nevertheless, the selection of tokens to retain remains controlled by the sampling strategy rather than by an explicitly learned retention policy.
% \subsubsection{Sampling}
\paragraph*{Sampling}
% Unlike training, which reconstructs $\bm Y_0$ from an independently sampled $\bm F$ in a single step, inference applies the same model iteratively. 
Although GRN is trained with the one-step objective in eq.~\ref{eq:grn_objective} to reconstruct $\bm Y_0$ from a randomly corrupted token map $\bm F$, inference applies the same model iteratively. 
% In contrast to the training input $\bm F$ in sec.~\ref{sec:grn_training}, 
$\bm F_k$ denotes the input token map at refinement step $k$. Starting from a uniformly sampled target token map $\bm F_0=\bm Y_{\mathrm{rand}}$, GRN predicts $\widehat{\bm Y}_{k+1}$ from $\bm F_k$ at refinement step $k$. A binary selection map $\bm S_{k+1}$ is sampled using retention
ratio $l_{k+1}$ to construct the next input:
\begin{equation}
\begin{aligned}
\bm F_{k+1}
&=
\bm S_{k+1}\odot\widehat{\bm Y}_{k+1}
+
(\bm 1-\bm S_{k+1})\odot\bm Y_{\mathrm{rand}}, \\
l_{k+1}
&=
\rho\left[
1-\cos\left(\frac{\pi}{2}\tau_{k+1}\right)
\right],
\end{aligned}
\label{eq:grn_inference}
\end{equation}
where $\rho=0.999$ and
$\tau_{k+1}=\min\{(k+1)/(K-1),1\}$ denotes the normalized refinement progress over $K$ steps. Predictions not selected by $\bm S_{k+1}$ are replaced by $\bm Y_{\mathrm{rand}}$ and re-predicted at the next step, allowing earlier errors to be corrected. After $K$ refinement steps, the final prediction $\widehat{\bm Y}_K$ is returned.

% and
% updates the input token map as follows:
% \begin{equation}
% \begin{aligned}
% l_{k+1}
% &=
% \rho\left[
% 1-\cos\left(\frac{\pi}{2}\tau_{k+1}\right)
% \right],\\
% S_{k+1}^{(i,j)}
% &\sim\operatorname{Bernoulli}(l_{k+1}),\\
% \bm F_{k+1}
% &=
% \bm S_{k+1}\odot\widehat{\bm Y}_{k+1}
% +
% (\bm 1-\bm S_{k+1})\odot\bm Y_{\mathrm{rand}}.
% \end{aligned}
% \label{eq:grn_inference}
% \end{equation}
% where $\tau_{k+1}=(k+1)/K$ denotes the normalized refinement progress,
% $K$ is the number of refinement steps, and $\rho$ is the maximum
% retention ratio.

% During inference, the retention ratio $l_t$ is determined by the refinement step $t$ using the cosine schedule.
% \begin{equation}
% \begin{aligned}
% \bm{F}_{t+1}
% &=
% \bm{S}_{t+1}\odot\widehat{\bm{Y}}_{t+1}
% +
% (1-\bm{S}_{t+1})\odot\bm{R},\\
% l_{t+1} &=
% \rho\left[
% \bm 1-\cos\left(\frac{\pi}{2}\tau_{t+1}\right)
% \right],
% \quad
% S_{t+1,i}\sim\operatorname{Bernoulli}(l_{t+1})
% \end{aligned}
% \label{eq:grn_inference}
% \end{equation}
\section{Related Works}
% \subsection{Autoregressive and Non-autoregressive TTS}
% \subsection{Neural Speech Tokenizer}
\subsection{Masked diffusion models for TTS}
Masked diffusion models (MDMs)~\cite{austin2021structured} extend diffusion modeling to discrete data spaces and have been applied to image~\cite{Chang2022MaskGITMG} and text generation~\cite{wang2026revolutionizing, nie2026large}. Unlike continuous diffusion models that recover data by progressively denoising Gaussian noise, MDMs define the denoising process by revealing masked tokens. Although the denoising process remains sequential, inference latency depends on the number of denoising steps rather than on the output sequence length in AR sequence modeling. 

MaskGIT~\cite{Chang2022MaskGITMG} is a foundational MDM-based work for image generation, which utilizes a bidirectional Transformer to predict masked tokens and determines a subset of high-confidence predictions to keep at each iteration. 
MaskGCT~\cite{wang2025maskgct} extends this paradigm to the zero-shot text-to-speech domain through text-to-semantic and semantic-to-acoustic generation, implemented as two MDM-based modules in a two-stage cascaded system.
Earlier two-stage TTS systems~\cite{Borsos2023SoundStormEP} applied MDMs only to the semantic-to-acoustic module, while retaining an AR text-to-semantic module that can still lead to unreliable and high latency generation.
However the accumulation of errors caused by eq.~\ref{eq:mdm_denoising} have been shown to degrade the quality of audio generation. 
To alleviate this issue, subsequent work~\cite{gallego2025single} simplified the original two-stage pipeline into a single-stage architecture using a bidirectional Transformer~\cite{Vaswani2017AttentionIA}. 
OmniVoice~\cite{zhu2026omnivoice} further formulated zero-shot TTS as a single-stage MDM that maps text directly to acoustic tokens.
It was initialized from a pre-trained AR large language model (LLM)~\cite{qwen3technicalreport} and adopted full-codebook random masking to improve model effectiveness, training efficiency, and speech intelligibility.

\subsection{Accumulationg errors analysis for masked diffusion models}
% Theoretically, though MDM-based TTS models can finishing inference in one go, however an AR inference decoding is practically utilized to balance the speech quality and latency for several steps.
% Thus, MDMs-based TTS models still suffer from error propagation during inference, which is similar to the exposure bias in AR modelling.
% Concretely, the limitation remains that, tokens can not be modified after its unmaskings, even if errors are introduced. 
% To alleviate the error accumulation, some error corrections has been explored for MDMs~\cite{schiff2026learn, wang2026remasking, zhao2026informed} and MDMs related applications, e.g., visual synthesis~\cite{lezama2022discrete, han2026generative}. ReMDM~\cite{wang2026remasking} revise tokens generated from prediction steps by additional correction steps, which is a training-free self-correction mechanism.
% Other works~\cite{huang2026don, kim2026finetuning, lezama2022discrete, schiff2026learn} are trained to either predict the positions of incorrect tokens or incorrect token themselves using an additional head or modules.
Although MDM-based TTS models can theoretically complete inference in a single step, multi-step iterative decoding is typically used in practice to balance speech quality and latency.
Thus, MDM-based TTS models still suffer from error propagation during inference, similar to exposure bias in AR modeling.
Specifically, tokens cannot be modified after they are unmasked, even if errors have been introduced. 
To alleviate error accumulation, several error-correction methods have been explored for MDMs~\cite{schiff2026learn, wang2026remasking, zhao2026informed} and related applications, such as visual synthesis~\cite{lezama2022discrete, han2026generative}. ReMDM~\cite{wang2026remasking} revises tokens generated during prediction through additional correction steps, providing a training-free self-correction mechanism.
Other methods~\cite{huang2026don, kim2026finetuning, lezama2022discrete, schiff2026learn} are trained to predict either the positions of incorrect tokens or the correct tokens themselves using additional heads or modules.
In contrast to GRN~\cite{han2026generative} and Progressive Self-Correction~\cite{schiff2026learn}, no such error correction investigation has been conducted in TTS and this is what this paper is aiming to accomplish.
\section{Experiments}
\subsection{Experimental setup}
% \subsubsection{Dataset}
% \noindent\textbf{Dataset} 
\paragraph*{Dataset}
For the ablation studies, we train on LibriTTS-R~\cite{Koizumi2023LibriTTSRAR} (585 hours of English at 24 kHz) and the premium subset of WenetSpeech4TTS~\cite{Ma2024WenetSpeech4TTSA1} (945 hours of Mandarin at 16 kHz). The primary model is trained on the English and Mandarin subsets of Emilia~\cite{he2024emilia}, containing 46.8k and 49.9k hours of speech, respectively.
% LibriTTS-R is a restored version of LibriTTS containing approximately 585 hours of 24-kHz English speech from 2,456 speakers; it preserves the original utterances while improving their recording quality~\cite{koizumi2023librittsr}. WenetSpeech4TTS is a multi-domain Mandarin corpus refined from WenetSpeech through improved segmentation, enhancement, transcription, and speaker filtering. We use its Premium subset, which contains approximately 945 hours of the highest-quality speech selected by DNSMOS scores~\cite{ma2024wenetspeech4tts}.

% For final-model training, we use the English and Chinese subsets of Emilia, a large-scale multilingual corpus collected from diverse in-the-wild sources and covering a broad range of speakers, domains, and speaking styles~\cite{he2024emilia}. 
We evaluate Revise-TTS on the LibriSpeech-PC~\cite{meister2023librispeechpc} test-clean set~\cite{chen2025f5} and the bilingual benchmark introduced by Seed-TTS~\cite{anastassiou2024seedtts}. 
% In total, the evaluation sets contain 2,215 English samples and 2,020 Mandarin samples.
For objective evaluation, these testing sets contain 2,215 English samples and 2,020 Mandarin samples in total.
% \subsubsection{Implementation}
\paragraph*{Implementation}
% Revise-TTS is built following OmniVoice~\cite{zhu2026omnivoice} with GRN framework, which can be initialized with the pre-trained LLM weights of Qwen3-0.6B~\cite{yang2025qwen3}.
% The speech tokenizer is impletmented using XCodec 2.0~\cite{Ye2025LlasaST} as the backbone with different audio quantizaiton algorithm, e.g., Finite Scalar uantization (FSQ)~\cite{mentzer2024finite} or Hierarchical Binary Quantization (HBQ)~\cite{han2026generative}. 
Revise-TTS follows OmniVoice~\cite{zhu2026omnivoice} as the backbone and uses an XCodec 2.0~\cite{Ye2025LlasaST} speech tokenizer with either FSQ~\cite{mentzer2024finite} or HBQ~\cite{han2026grn} quantization.
% \subsubsection{Training and inference}
\paragraph*{Training and inference}
For the ablation study and primary model, we train on 4 and 16 NVIDIA A100 GPUs with batch sizes of 1.8k and 6.5k seconds (20,480 tokens per GPU) for 100k and 250k steps (one epoch), respectively. We use the AdamW optimizer with $\beta_1=0.9$, $\beta_2=0.999$, $\epsilon=10^{-8}$, and a learning rate of $10^{-4}$. Inference uses 32 iterative refinement steps.

% \subsubsection{Metrics}
\paragraph*{Metrics}
Objective metrics include speaker similarity, intelligibility, and naturalness. Speaker similarity is the cosine similarity between WavLM-based embeddings~\cite{chen2022wavlm} of ground-truth and synthesized speech. We measure intelligibility using WER for English and CER for Mandarin (both denoted WER below), with a HuBERT-based ASR model~\cite{hsu2021hubert} for LibriSpeech-PC test-clean, Paraformer-zh~\cite{gao2022paraformer} for Mandarin Seed-TTS, and Whisper~large-v3~\cite{radford2023whisper} for English Seed-TTS. Naturalness is measured using UTMOS~\cite{saeki2022utmos}.

Subjective evaluation uses CMOS ($[-3,3]$) for relative speech-quality preference and SMOS ($[0,5]$) for perceived speaker similarity to the prompt audio. We randomly select 10, 10, and 20 audio samples from LibriSpeech-PC, Seed-TTS EN, and Seed-TTS ZH, respectively, to form the English and Mandarin test sets. Each sample is evaluated by at least 15 native speakers via Amazon Mechanical Turk.
% \subsection{Experimental Results}
% \subsection{Initialization with Qwen}
% Table~\ref{tab:qwen_initialization} compares initialization from Qwen3-0.6B with random initialization under different learning rates using the ablation-study training data. All results are obtained after 100K TTS training steps. For the directly comparable learning rate of $1\times10^{-4}$, Qwen initialization consistently reduces WER from 4.76 to 2.33 on Libri, from 15.65 to 3.20 on Seed-en, and from 3.22 to 2.18 on Seed-zh. The particularly large improvement on Seed-en suggests that pretrained linguistic representations provide a substantially stronger initialization and improve intelligibility across all three benchmarks.
% \begin{table}[t]
% \centering
% \caption{WER of Initialization with Qwen}
% \label{tab:qwen_initialization}
% \small
% \setlength{\tabcolsep}{5pt}
% \begin{tabular}{lcccc}
% \toprule
% \textbf{Model} & \textbf{lr} &
% \textbf{Libri} $\downarrow$ &
% \textbf{Seed-en} $\downarrow$ &
% \textbf{Seed-zh} $\downarrow$ \\
% \midrule
% Qwen initialization & 1e-4 & 2.33 & 3.20 & 2.18 \\
% \midrule
% \multirow{4}{*}{Random initialization} & 1e-4 & 4.76 & 15.65 & 3.22 \\
%  & 5e-5 &  &  &  \\
%  & 5e-4 &  &  &  \\
%  & 1e-3 &  &  &  \\
% \bottomrule
% \end{tabular}
% \end{table}

\subsection{Ablation study}
Table~\ref{tab:FSQ_GRN_ind} compares between FSQ and $\mathrm{GRN}_{\mathrm{ind}}$ with different HBQ configurations. HBQ $[2^2]^8$ achieves the lowest WER on all three testing sets and the highest UTMOS on LibriSPeech-PC and Seed-TTS en. In contrast, HBQ $[2^4]^8$ achieves the highest SIM on all three testing sets. These results show that increasing either $C$ or $M$ does not lead to monotonic improvements. Instead, different configurations exhibit trade-offs among intelligibility, speaker similarity, and naturalness. At the same codebook size of $2^{16}$, HBQ $[2^2]^8$ consistently outperforms FSQ $[4]^8$ across all metrics and testing sets. The two FSQ configurations also produce comprehensively worse performance than the HBQ configurations.

To examine the effectiveness of the prediction modes, we compare $\mathrm{GRN}_{\mathrm{ind}}$ and $\mathrm{GRN}_{\mathrm{bit}}$ under identical $(C,M)$ configurations in Table~\ref{tab:hbq_index_bit_100k}. For $(C,M)=(8,2)$, 
% $\mathrm{GRN}_{\mathrm{ind}}$ consistently outperforms
% $\mathrm{GRN}_{\mathrm{bit}}$ across all metrics and testing sets.
% As $M$ increases to four or six, $\mathrm{GRN}_{\mathrm{bit}}$ becomes more competitive and achieves lower WER on LS-PC and Seed-en. 
% However, $\mathrm{GRN}_{\mathrm{ind}}$ retains lower WER on Seed-zh and generally higher UTMOS. 
although the two modes preserve the same information capacity, they remains the difference of classifications discussed in sec.\ref{sec:HBQ}.
% result in different prediction problems: $\mathrm{GRN}_{\mathrm{ind}}$ performs $C$ parallel $2^M$-way classifications, whereas $\mathrm{GRN}_{\mathrm{bit}}$ decomposes them into $CM$ binary classifications. 
Bit prediction therefore reduces the output space of classifications as $M$ increases, but does not provide uniform gains across all evaluation metrics.

\begin{table}[t]
\centering
\caption{Comparison between FSQ and $\mathrm{GRN}_{\mathrm{ind}}$ with different configurations. 
% Objective evaluation results of XCodec models trained on the small dataset, using 130K-step codec checkpoints and 100K-step TTS checkpoints. LS-PC denotes LibriSpeech-PC test-clean, while Seed-en and Seed-zh denote the English and Mandarin Seed-TTS testing sets, respectively. 
Best and second best are shown in bold and underline.}
\label{tab:FSQ_GRN_ind}
\small
\setlength{\tabcolsep}{3pt}
\renewcommand{\arraystretch}{0.82}
\begin{tabular}{@{}llccc@{}}
\toprule
\textbf{Quantization} & \textbf{Testing set} &
\textbf{SIM} $\uparrow$ & \textbf{WER} $\downarrow$ &
\textbf{UTMOS} $\uparrow$ \\
\midrule
\multirow{3}{*}{HBQ $[2^2]^8$}
& LS-PC   & 0.427 & \textbf{1.86} & \textbf{4.15} \\
& Seed-en & \underline{0.406} & \textbf{5.06} & \textbf{3.99} \\
& Seed-zh & 0.576 & \textbf{1.83} & 2.78 \\
\cmidrule(lr){1-5}
\multirow{3}{*}{HBQ $[2^2]^{16}$}
& LS-PC   & 0.422 & 2.37 & 3.85 \\
& Seed-en & 0.395 & 9.74 & 3.43 \\
& Seed-zh & \underline{0.598} & 2.71 & 2.46 \\
\cmidrule(lr){1-5}
\multirow{3}{*}{HBQ $[2^4]^4$}
& LS-PC   & 0.384 & \underline{1.94} & \underline{4.09} \\
& Seed-en & 0.369 & 8.59 & \underline{3.92} \\
& Seed-zh & 0.538 & 2.70 & \underline{2.83} \\
\cmidrule(lr){1-5}
\multirow{3}{*}{HBQ $[2^4]^8$}
& LS-PC   & \textbf{0.444} & 2.29 & 4.03 \\
& Seed-en & \textbf{0.419} & 7.17 & 3.70 \\
& Seed-zh & \textbf{0.607} & 2.44 & 2.67 \\
\cmidrule(lr){1-5}
\multirow{3}{*}{HBQ $[2^6]^8$}
& LS-PC   & \underline{0.432} & 2.23 & 4.08 \\
& Seed-en & 0.385 & 6.38 & 3.88 \\
& Seed-zh & 0.585 & \underline{2.37} & \textbf{2.88} \\
\midrule
\multirow{3}{*}{FSQ $[4]^8$}
& LS-PC   & 0.290 & 2.07 & 4.04 \\
& Seed-en & 0.270 & \underline{5.93} & 3.71 \\
& Seed-zh & 0.514 & 2.60 & 2.36 \\
\cmidrule(lr){1-5}
\multirow{3}{*}{FSQ $[8,5,5,5]$}
& LS-PC   & 0.259 & 3.08 & 3.71 \\
& Seed-en & 0.236 & 6.94 & 3.74 \\
& Seed-zh & 0.426 & 3.65 & 2.70 \\
\bottomrule
\end{tabular}
\end{table}

% Figure~\ref{fig:small_dataset_ablation_study} compares FSQ with HBQ using two, four, and six quantization rounds on the ablation-study data. With the same $2^{16}$ codebook size, HBQ-R2 substantially outperforms FSQ in SIM-o and UTMOS and achieves lower final WER across all three benchmarks, indicating that the improvement arises from the quantization structure rather than codebook size alone. Among the HBQ variants, R2 converges fastest and provides the best intelligibility, R4 achieves the highest speaker similarity, and R6 obtains the best UTMOS on Seed-TTS test-zh. Increasing the number of rounds therefore does not yield uniform gains, revealing a trade-off among intelligibility, speaker similarity, and naturalness.

\begin{table}[htbp]
\centering
\caption{Comparison between $\mathrm{GRN}_{\mathrm{ind}}$ and $\mathrm{GRN}_{\mathrm{bit}}$.}
\label{tab:hbq_index_bit_100k}
\small
\setlength{\tabcolsep}{2pt}
\renewcommand{\arraystretch}{0.82}
\begin{tabular}{@{}cllccc@{}}
\toprule
\textbf{Config.} & \textbf{Mode} & \textbf{Testing set} &
\textbf{SIM} $\uparrow$ & \textbf{WER} $\downarrow$ &
\textbf{UTMOS} $\uparrow$ \\
\midrule
\multirow{6}{*}{$[2^2]^8$}
& \multirow{3}{*}{GRN-ind}
& LS-PC   & 0.427 & \textbf{1.86} & \textbf{4.15} \\
& & Seed-en & 0.406 & \textbf{5.06} & \textbf{3.99} \\
& & Seed-zh & 0.576 & \textbf{1.83} & \underline{2.78} \\
\cmidrule(lr){2-6}
& \multirow{3}{*}{GRN-bit}
& LS-PC   & 0.382 & 2.15 & 3.94 \\
& & Seed-en & 0.366 & 12.71 & 3.73 \\
& & Seed-zh & 0.542 & 2.74 & 2.53 \\
\midrule
\multirow{6}{*}{$[2^4]^8$}
& \multirow{3}{*}{GRN-ind}
& LS-PC   & \textbf{0.444} & 2.29 & 4.03 \\
& & Seed-en & \textbf{0.419} & 7.17 & 3.70 \\
& & Seed-zh & \textbf{0.607} & 2.44 & 2.67 \\
\cmidrule(lr){2-6}
& \multirow{3}{*}{GRN-bit}
& LS-PC   & \textbf{0.444} & \underline{2.06} & 4.03 \\
& & Seed-en & \underline{0.411} & 6.96 & 3.72 \\
& & Seed-zh & \underline{0.600} & 2.69 & 2.63 \\
% \midrule
% \multirow{6}{*}{$[2^6]^8$}
% & \multirow{3}{*}{GRN-ind}
% & LS-PC   & \underline{0.432} & 2.23 & \underline{4.08} \\
% & & Seed-en & 0.385 & 6.38 & \underline{3.88} \\
% & & Seed-zh & 0.585 & \underline{2.37} & \textbf{2.88} \\
% \cmidrule(lr){2-6}
% & \multirow{3}{*}{GRN-bit}
% & LS-PC   & \underline{0.432} & 2.10 & 4.00 \\
% & & Seed-en & 0.406 & \underline{5.90} & 3.76 \\
% & & Seed-zh & 0.596 & 2.87 & 2.73 \\
\bottomrule
\end{tabular}
\end{table}

\subsection{Codec reconstruction}
% The configuration HBQ $[2^2]^8$ is training  Table~\ref{tab:codec_reconstruction_results} summarises the reconstruction performance the configuration HBQ $[2^2]^8$ training on Emilia. Higgs achieves the best performance on all metrics. Though official XCodec 2.0 was trained on larger dataset and more iterations than XCodec 2.0 with HBQ, XCodec 2.0 with HBQ quantization algorithm still achieves comparable performance.
The primary codec uses the HBQ $[2^2]^8$ configuration and is trained on Emilia. Table~\ref{tab:codec_reconstruction_results} summarizes its reconstruction performance. Higgs achieves the best on all metrics. Although the official XCodec 2.0 was trained on more data and for more iterations than XCodec 2.0 wth HBQ, the latter remains competitive.
\begin{table}[htbp]
\centering
\caption{Comparison between codecs: Higgs, official checkpoint of XCodec 2.0 and XCodec 2.0 with HBQ.}
\label{tab:codec_reconstruction_results}
\small
\setlength{\tabcolsep}{2.5pt}
\renewcommand{\arraystretch}{0.88}
\begin{tabular}{@{}lcccccc@{}}
\toprule
\multirow{2}{*}{\textbf{Codec}} &
\multicolumn{2}{c}{\textbf{LS-PC}} &
\multicolumn{2}{c}{\textbf{Seed-en}} &
\multicolumn{2}{c}{\textbf{Seed-zh}} \\
\cmidrule(lr){2-3}
\cmidrule(lr){4-5}
\cmidrule(lr){6-7}
& SIM $\uparrow$ & WER $\downarrow$
& SIM $\uparrow$ & WER $\downarrow$
& SIM $\uparrow$ & WER $\downarrow$ \\
\midrule
Higgs
& \textbf{0.877} & \textbf{2.33}
& \textbf{0.904} & \textbf{2.33}
& \textbf{0.903} & \textbf{1.20} \\
XCodec 2.0 + FSQ
& 0.781 & 3.08
& 0.821 & 2.80
& 0.840 & 1.44 \\
XCodec 2.0 + HBQ
& 0.764 & 3.60
& 0.802 & 3.12
& 0.830 & 1.92 \\
\bottomrule
\end{tabular}
\end{table}

\subsection{GRN fine-tuning}
Table~\ref{tab:fine-tuning} compares two fine-tuning modes: masking and GRN decoding. For a fair comparison, both use Higgs and the same training configuration, with 100k iterations on eight GPUs. The official OmniVoice checkpoint outperforms both fine-tuned models. GRN performs slightly worse than masking, particularly in speaker similarity across all three testing sets. This gap may arise because OmniVoice uses reinforcement learning (RL) to improve speaker similarity. Nevertheless, the results demonstrate that GRN can be easily extended to other MDM-based TTS systems.
\begin{table}[htbpt]
\centering
\caption{Comparison of different decoding modes using the Higgs codec,
fine-tuned on Emilia EN/ZH with 8 GPUs.}
\label{tab:fine-tuning}
\small
\setlength{\tabcolsep}{1pt}
\renewcommand{\arraystretch}{0.88}
\begin{tabular}{@{}lccccccc@{}}
\toprule
\multirow{2}{*}{\textbf{Mode}} &
\multirow{2}{*}{\textbf{Steps}} &
\multicolumn{2}{c}{\textbf{LS-PC}} &
\multicolumn{2}{c}{\textbf{Seed-en}} &
\multicolumn{2}{c}{\textbf{Seed-zh}} \\
\cmidrule(lr){3-4}
\cmidrule(lr){5-6}
\cmidrule(lr){7-8}
& & SIM $\uparrow$ & WER $\downarrow$
& SIM $\uparrow$ & WER $\downarrow$
& SIM $\uparrow$ & WER $\downarrow$ \\
\midrule
Masking & 100k
& 0.689 & 1.81
& 0.701 & 1.98
& 0.762 & 1.06 \\
GRN & 100k
& 0.646 & 1.80
& 0.637 & 2.50
& 0.719 & 1.27 \\
\midrule
OmniVoice & --
& \textbf{0.697} & \textbf{1.57}
& \textbf{0.717} & \textbf{1.72}
& \textbf{0.765} & \textbf{0.89} \\
\bottomrule
\end{tabular}
\end{table}

\subsection{Objective evaluation and subjective evaluation}
% Table~\ref{tab:objective_evaluation} compares ReviseTTS with SOTA NAR TTS models. 
% % Although Revise-TTS achieves relative low similarity score, we argue it may improve using the configuration of HBQ $[2^4]^8$ or through RL post-training. 
% ReviseTTS demonstrates strong Mandarin intelligibility, achieving a WER of 1.19\%, and surpasses the ground truth in naturalness across all three test subsets. However, these gains are accompanied by lower speaker similarity, which may potentially be improved with the HBQ $[2^4]^8$ configuration or RL post-training.

Table~\ref{tab:objective_evaluation} compares ReviseTTS with state-of-the-art NAR TTS systems. ReviseTTS demonstrates strong Mandarin intelligibility, achieving a WER of 1.19\%, and obtains slightly higher UTMOS scores than the ground truth across all three test sets. However, these results come at the cost of lower speaker similarity, which may be improved by adopting the HBQ $[2^4]^8$ configuration or RL-based post-training.
In the subjective evaluation, OmniVoice achieves positive CMOS scores on both the English and Mandarin test sets, whereas ReviseTTS obtains comparable perceived quality to F5-TTS and MaskGCT, respectively.
\begin{table}[htbp]
\centering
\caption{Objective and subjective evaluation results.}
\label{tab:objective_evaluation}
\small
\setlength{\tabcolsep}{3pt}
\renewcommand{\arraystretch}{0.82}
\begin{tabular}{@{}llcccc@{}}
\toprule
\textbf{Model} & \textbf{Test set} &
\textbf{SIM} $\uparrow$ &
\textbf{WER} $\downarrow$ &
\textbf{UTMOS} $\uparrow$ &
\textbf{CMOS} $\uparrow$ \\
\midrule

\multirow{3}{*}{GT}
& LS-PC   & 0.690 & \underline{1.87\%} & 4.01
& \multirow{2}{*}{0.00} \\
& Seed-en & \textbf{0.734} & 2.14\% & 3.52 & \\
& Seed-zh & 0.755 & 1.25\% & 2.78
& 0.00 \\

\cmidrule(lr){1-6}
\multirow{3}{*}{F5-TTS}
& LS-PC   & 0.655 & 1.89\% & 3.89
& \multirow{2}{*}{$-0.30 \pm 1.69$} \\
& Seed-en & 0.664 & \underline{1.85\%} & \underline{3.72} & \\
& Seed-zh & 0.750 & 1.53\% & \underline{2.93} & $-0.47 \pm 1.81$ \\

\cmidrule(lr){1-6}
\multirow{3}{*}{MaskGCT}
& LS-PC   & \underline{0.691} & 2.26\% & 3.91
& \multirow{2}{*}{$-0.21 \pm 1.70$} \\
& Seed-en & 0.713 & 2.88\% & 3.55 & \\
& Seed-zh & \textbf{0.773} & 2.40\% & 2.63 & \underline{$-0.37 \pm 1.73$} \\

\cmidrule(lr){1-6}
\multirow{3}{*}{OmniVoice}
& LS-PC   & \textbf{0.697} & \textbf{1.57\%} & \textbf{4.23}
& \multirow{2}{*}{$\textbf{+0.28} \pm 1.71$} \\
& Seed-en & \underline{0.717} & \textbf{1.72\%} & \textbf{3.88} & \\
& Seed-zh & \underline{0.765} & \textbf{0.89\%} & \textbf{3.05}
& {$\textbf{+0.13} \pm 1.93$} \\

\midrule
\multirow{3}{*}{Revise-TTS}
& LS-PC   & 0.519 & 2.00\% & \underline{4.03}
& \multirow{2}{*}{$-0.34 \pm 1.67$} \\
& Seed-en & 0.532 & 2.91\% & 3.60 & \\
& Seed-zh & 0.645 & \underline{1.19\%} & 2.80
& $-0.63 \pm 1.88$ \\

\bottomrule
\end{tabular}
\end{table}

% \begin{table}[t]
% \centering
% \caption{Subjective evaluation.}
% \label{tab:subjective_evaluation}
% \small
% \setlength{\tabcolsep}{8pt}
% \renewcommand{\arraystretch}{0.88}
% \begin{tabular}{@{}lcc@{}}
% \toprule
% \textbf{Model} & CMOS $\uparrow$ & SMOS $\uparrow$ \\
% \midrule
% Grount truth & & \\
% F5-TTS     & & \\
% MaskGCT    & & \\
% OmniVoice  & & \\
% \midrule
% Revise-TTS & & \\
% \bottomrule
% \end{tabular}
% \end{table}

\section{Conclusion}
% This paper proposed Revise-TTS, a new framework called GRN to solve the issue of error accumulation in MDM-based TTS. The novelty of Revise-TTS is to introduce non-masking training and overlapping inference, where a small proportion of tokens would be revised through iteratively refinement. To better adjust the GRN framework, hierarchical binary quantization is proposed to provide natural bit-wise prediction and explicit coarse-to-fine structure of token maps.
This paper proposed Revise-TTS, a single-stage zero-shot TTS framework that applies GRN non-masking training and overlapping inference to MDM-based TTS. During training, the model reconstructs the token maps from mixtures of ground-truth and random tokens. During inference, it iteratively revises a small proportion of previous predictions. Revise-TTS also adopts HBQ to provide an explicit coarse-to-fine token structure and support explicit bit-wise prediction. Experiments show that HBQ outperforms FSQ under matched codebook. Revise-TTS achieves competitive intelligibility and naturalness, although speaker similarity remains improvement.
\clearpage
\bibliographystyle{IEEEbib}
\bibliography{strings,refs}

\end{document}